\documentclass[twocolumn,pra,superscriptaddress,showpacs]{revtex4}
\usepackage{graphicx}
\usepackage{dcolumn}
\usepackage{amsmath}
\usepackage{bm}
\usepackage{color}

\begin{document}

\title{Laser Intensity Dependence of Photoassociation in Ultracold Metastable Helium}
\author{Daniel G. Cocks}
\affiliation{School of Mathematics, Physics and Information Technology,
James Cook University, Townsville, Australia 4811}
\author{Ian B. Whittingham}
\affiliation{School of Mathematics, Physics and Information Technology,
James Cook University, Townsville, Australia 4811}

\date{\today}

\begin{abstract}
Photoassociation of spin-polarized metastable helium to the three lowest rovibrational 
levels of the $J=1$, $0_u^+$ state asymptoting to 2$s\,{}^{3}$S$_{1}+$2$p\,{}^{3}$P$_{0}$ 
is studied using a second-order perturbative treatment of the line shifts valid for low
laser intensities, and two variants of a non-perturbative close-coupled treatment, 
one based upon dressed states of the matter plus laser system,
and the other on a modified radiative coupling which vanishes asymptotically, thus 
simulating experimental conditions. These non-perturbative treatments are valid for arbitrary
laser intensities and yield the complete photoassociation resonance profile. Both variants
give nearly identical results for the line shifts and widths of the resonances and show that 
their dependence upon laser intensity is very close to linear and quadratic respectively
for the two lowest levels.
The resonance profiles are superimposed upon a significant background loss, a feature for 
this metastable helium system not present in studies of photoassociation in other systems, 
which is due to the very shallow nature of the excited state $0_u^+$ potential. 
The results for the line shifts from the close-coupled and perturbative
calculations agree very closely at low laser intensities. 
\end{abstract}

\maketitle

\section{Introduction}

Photoassociation (PA) experiments in which two interacting ultracold atoms (usually 
ground state alkali atoms or metastable rare gas atoms) are resonantly excited by
a laser to a molecular bound state provide a powerful technique to study
the dynamics of ultracold collisions \cite{Thors87,Weiner99}. Since the colliding atoms
are so cold, the energy of the initial scattering state is well determined and
the resonant laser energies corresponding to transitions to various bound states
produce a very high resolution spectrum ($< 1$ MHz). At this level of precision, 
energy level shifts induced by the PA laser can be significant 
\cite{Nashifts,Lishifts,Portier06}
and an understanding of the dependence of the energy level shifts upon the laser
intensity, polarization and frequency is crucial.

The short lived bound states created in photoassociation are unique
in that they can occupy both the small interatomic regions of conventional
molecules and the unusual large interatomic regions of hundreds of
Bohr radii depending upon the particular intermolecular potential.
For purely long range molecules, the interaction between
the atoms, and hence the molecular potentials, arises from dispersion forces
that depend mainly upon well known atomic parameters.
Extensions of the photoassociation technique can
be used to determine the lifetimes of excited states \cite{Dall08}, scattering lengths
of ground states \cite{Julienne96}, create ground state cold molecules
\cite{Fioretti98}
and drive other laser orientated processes.

Although ultracold physics and the achievement of Bose-Einstein condensation
has its roots in alkali species, the cooling of excited state species
such as metastable helium opens many more opportunities for experiment.
Metastable rare gas atoms offer exciting experimental detection
strategies to study quantum gases as their large internal energy can be released
through Penning and associative ionization during interatomic collisions
and through ejection of electrons when the atoms strike a metal surface.
Additionally, trap purity is more easily maintained in a metastable gas, as rare
gas ground states are not trapped by the same mechanisms as metastable states.
The metastable atoms are generally spin-polarized in order to suppress the
autoionization rate and to thereby attain large numbers of trapped atoms.
As well, no hyperfine structure is present in rare gas species, making
them more desirable to investigate than many other species.

In metastable helium a number of experimental investigations have
been conducted using photoassociation as the diagnostic tool. 
Hershbach \textit{et al.} \cite{Hersch00} observed bound states that 
dissociate to the 2$s\,{}^{3}$S$_{1}+$2$p\,{}^{3}$P$_{2}$ atomic limit, 
L\'{e}onard \textit{et al.} \cite{Leonard03} studied some purely long-range 
bound states with binding energies $\leq $ 1.43 GHz, dissociating to the
2$s\,{}^{3}$S$_{1}+$2$p\,{}^{3}$P$_{0}$ limit, Kim \textit{et al.} 
\cite{Kim04} and van Rijnbach \cite{Rijn04} observed detailed structure of over 40
peaks associated with bound states with binding energies $\leq $~13.57 GHz that 
dissociate to the 2$s\,{}^{3}$S$_{1}+$2$p\,{}^{3}$P$_{2}$ limit, and van Rijnbach 
\cite{Rijn04} has observed six peaks lying within 0.5 GHz of the  
2$s\,{}^{3}$S$_{1}+$2$p\,{}^{3}$P$_{1}$ limit. 
Theoretical analyses of the long-range bound states dissociating to the
2$s\,{}^{3}$S$_{1}+$2$p\,{}^{3}$P$_{0}$ limit have been completed using a single 
channel adiabatic calculation \cite{Leonard04} and full multichannel 
calculations \cite{Venturi03}. Both calculations use retarded long-range 
Born-Oppenheimer dispersion potentials  and give excellent agreement with the 
measured binding energies.  
Most of the 40 peaks associated with the 2$s\,{}^{3}$S$_{1}+$2$p\,{}^{3}$P$_{2}$ 
limit were identified \cite{DGL05} using the accumulated phase technique 
for a single channel calculation of the bound states supported by a hybrid 
quintet potential constructed from short-range \textit{ab initio} 
${}^{5}\Sigma^{+}_{g/u}$ and ${}^{5}\Pi^{+}_{g/u}$  potentials matched onto 
long-range retarded dispersion potentials. Recently Deguilhem \textit{et al.} 
\cite{DLGD09} have revisited 
the analysis of the PA peaks associated with the 2$s\,{}^{3}$S$_{1}+$2$p\,{}^{3}$P$_{1,2}$ 
limits using fully \textit{ab initio} short range potentials.
Light-induced level shifts of several vibrational states in the long-range 
$J=1,0^{+}_{u}$ potential associated with the  
2$s\,{}^{3}$S$_{1}+$2$p\,{}^{3}$P$_{0}$ asymptote have been measured by Kim 
\textit{et al.} \cite{Kim05} and two-photon photoassociation spectroscopy 
has recently been used \cite{Moal06} to accurately measure the binding energy 
of the least bound vibrational level
($v=14$) of the metastable ${}^{5}\Sigma^{+}_{g}$ state formed during the collision
of two spin-polarized metastable helium atoms. The measured binding energy 
$E_{b}(v=14) =91.35 \pm 0.06$ MHz,
combined with the new  \textit{ab initio} ${}^{5}\Sigma^{+}_{g}$ potential of
Przybytek and Jeziorski \cite{Przyb05}, yielded the high precision value
$a=7.512 \pm 0.005$ nm for the $s$-wave scattering length.

All previous theoretical investigations of photoassociation involve limiting 
assumptions that may not be valid in the present investigation. Perturbative 
treatments of the radiative coupling \cite{NWWJ94,BJ99} predict a linear
dependence of the line shifts upon laser intensity but are only valid for low
laser intensities. The analytical method of Simoni \textit{et al.} \cite{SJTW02}
is valid for arbitrary laser intensities but assumes the radiative coupling 
vanishes asymptotically, thus avoiding the use of dressed states for the atoms
in the radiation field. This is only valid when the coupling is negligible
compared to the detuning to the atomic transition. The most detailed treatment 
is that of Napolitano \cite{Napolitano98} which employs a full multichannel 
treatment using dressed $s$- and $d$-wave states. However fine structure is 
neglected and it is assumed that the laser detunings are large compared 
to the radiative coupling.

Theoretical investigations of photoassociation in ultracold metastable helium 
are limited to the perturbative analysis of the light-induced energy level
shifts of the excited $J=1,0^{+}_{u}$ rovibrational state \cite{Portier06}
based upon the theories of \cite{BJ99,SJTW02}. In this model the shifts 
are linear in the laser intensity and in the $s$-wave scattering length $a$. 
However, as the $J=1,0^{+}_{u}$ potential is very shallow, the laser detuning
is quite small and the validity of not using dressed states needs investigation 
\cite{Leduc05}.

\begin{figure}
\includegraphics[width=0.95\linewidth]{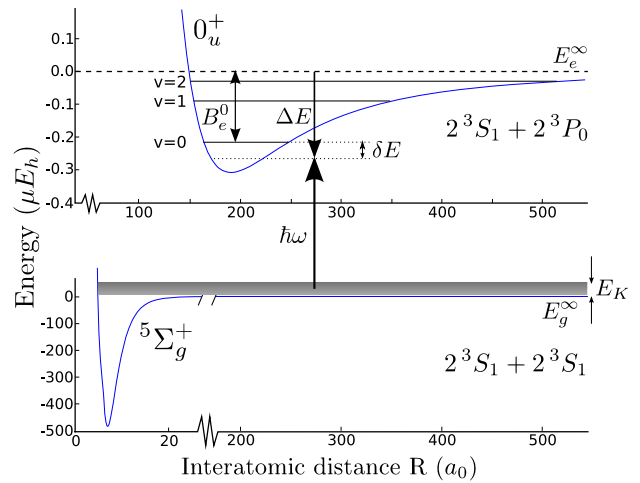}
\caption{\label{fig:PAdiag}
Photoassociation in metastable helium. Two spin-polarized metastable helium 
atoms ($2s\,^{3}$S$_{1}+2s\;^{3}$S$_{1}$) with very low kinetic energy $E_{K}$ 
absorb a photon of frequency $\omega $ and form a short-lived molecule in the 
rovibrational level $v$ of the $2s\,^{3}$S$_{1}+2p\;^{3}$P$_{0}$ $0_u^+ $ excited state. 
The molecule can then spontaneously decay back to the disassociated metastable state 
(all other decay paths are spin-forbidden). 
As illustrated here for the $v=0$ level, $\delta E$ is the energy level shift induced by 
the laser field, $\Delta E$ is the laser detuning energy from the separated atom resonance
and $B^{0}_{e}$ is the binding energy of the level.}
\end{figure}

The goal of this paper is to present a complete treatment
of photoassociation of spin-polarized metastable helium to the $J=1$, $0_u^+$
state asymptoting to 2$s\,{}^{3}$S$_{1}+$2$p\,{}^{3}$P$_{0}$. This process is illustrated 
in Fig. \ref{fig:PAdiag}. Of particular interest is the 
dependence of the photoassociation broadenings and level shifts upon laser intensity 
and polarization for small laser detunings comparable to the laser coupling. 

This paper is organised as follows. The general formalism of two atoms
colliding in the presence of a laser field is presented in section II, 
a perturbative treatment of laser-induced energy level shifts in section III 
and a non-perturbative close-coupled treatment of the photoassociation  
resonance profile in \text{section IV}. Section V presents our 
discussion and conclusions. Two appendices provide more details on the basis states 
used in our calculations and the evaluation of the matrix elements of the system
Hamiltonian in this basis. Atomic units are
used in the actual calculations, with lengths in Bohr radii $a_{0}$ and
energies in Hartree $E_{h}=\alpha ^{2}m_{e}c^{2}=27.211396$ eV.

\section{Two-Atom Collisions in a Light Field}

\subsection{Hamiltonian}

The total Hamiltonian for two atoms colliding in the presence of a radiation 
field is
\begin{equation}
\label{cw-1}
\hat{H} = \hat{H}_{\text{mol}} + \hat{H}_{\text{rad}} + \hat{H}_{\text{int}}
\end{equation}
where $\hat{H}_{\text{mol}}$ is the total molecular Hamiltonian (in barycentric
coordinates)
\begin{equation}
\label{cw-2}
\hat{H}_{\text{mol}} = \hat{T} + \hat{H}_{\text{rot}} + \hat{H}_{\text{el}}
+\hat{H}_{\text{fs}}.
\end{equation}
Here $\hat{T}=-(\hbar ^{2}/2\mu R)(\partial^{2}/\partial R^{2})\,R $ is the
kinetic energy operator, $\hat{H}_{\text{rot}} = \hat{l}^{2}/(2 \mu R^{2})$ is
the rotational operator for a system with relative angular momentum
$\hat{\bm{l}}$ and reduced mass $\mu $,
$\hat{H}_{\text{el}}=\hat{H}_{1}+\hat{H}_{2}+\hat{H}_{12}$ is the total
electronic Hamiltonian of the unperturbed atoms $\hat{H}_{1,2}$ and their 
electrostatic interaction $\hat{H}_{12}$, 
and $\hat{H}_{\text{fs}}$ describes the fine structure of the
atoms.  The Hamiltonian for the free radiation field is $\hat{H}_{\text{rad}}=
\sum_{\xi} \hbar \omega_{\xi}\hat{a}^{\dag}_{\xi}\hat{a}_{\xi}$ where
$\hat{a}^{\dag}_{\xi}\;(\hat{a}_{\xi})$ are the usual creation (annihilation) operators for
a photon of angular frequency $\omega_{\xi}$ and polarization
$\bm{\epsilon}_{\xi}$ so that the field states are
$|n,\omega_{\xi},\bm{\epsilon}_{\xi}\rangle =
(n!)^{-1/2}(\hat{a}^{\dag}_{\xi})^{n}|\text{vac} \rangle $.  The coupling between the
two atoms and the radiation field is
\begin{equation}
\label{cw-3a}
\hat{H}_{\text{int}} = -(\frac{e}{m})\sum_{i=1,2} \hat{\bm{p}}_{i}\cdot 
\hat{\bm{A}}(\bm{r}_{i})
\end{equation}
where $\hat{\bm{p}}_{i}=-i\hbar \bm{\nabla}_{\bm{r}_{i}}$. The vector potential is
\begin{equation}
\label{cw-3b}
\hat{\bm{A}}(\bm{r}_{i}) = \sum_{\xi}[\bm{\mathcal{E}}_{\xi}(\bm{r}_{i})\;\hat{a}_{\xi} + 
\bm{\mathcal{E}}_{\xi}(\bm{r}_{i})^{*}\;\hat{a}^{\dag}_{\xi}],
\end{equation}
where $\bm{\mathcal{E}}_{\xi}(\bm{r}_{i})=(\hbar /2 \omega_{\xi}\epsilon_{0}
\mathcal{V})^{1/2}\exp(i\bm{k}\cdot \bm{r}_{i})\bm{\epsilon}_{\xi}$ and 
$\mathcal{V}$ is the normalization volume. 
 
\subsection{Close-coupled equations}

The close-coupled equations describing the interaction of the two atoms in an
applied laser field with given angular frequency $\omega $ and given polarization
$\bm{\epsilon}_{\lambda}$ are obtained by expanding the energy eigenstates 
$|\Psi \rangle $ of $\hat{H}$ in terms of a basis of the form \cite{Mies81} 
\begin{equation}
\label{cw-4}
|\Phi_{a}, n\rangle \equiv |\Phi_{a}(R,q)\rangle \otimes |n, \omega , 
\bm{\epsilon}_{\lambda}\rangle 
\end{equation}
where $a$ denotes a set of approximate quantum numbers describing the 
electronic-rotational states of the molecule (to be discussed in Section IIC), 
and $q$ denotes the interatomic polar coordinates $(\theta ,\phi )$ and the 
electronic coordinates $(\bm{r}_{1},\bm{r}_{2})$. Using the expansion
\begin{equation}
\label{cw-5}
|\Psi \rangle = R^{-1} \left[ \sum_{g^{\prime}}G_{g^{\prime}}(R) 
|\Phi_{g^{\prime}},n \rangle + \sum_{e^{\prime}}G_{e^{\prime}}(R)
|\Phi_{e^{\prime}},n-1 \rangle \right],
\end{equation}
where $g^{\prime}(e^{\prime})$ labels the sets of metastable (excited) states, 
in $\hat{H}|\Psi \rangle = E |\Psi \rangle $ yields the close-coupled equations 
\begin{eqnarray}
\label{cw-6}
\sum_{g^{\prime}}\left[T^G_{g g^{\prime}}+V_{g g^{\prime}}
G_{g^{\prime}}(R)\right]
+ \sum_{e^{\prime}} V^{\text{int}}_{g e^{\prime}} G_{e^{\prime}}(R) \nonumber  \\
 =  (E-n\hbar \omega )G_{g}(R), \nonumber  \\
\sum_{e^{\prime}}\left[T^G_{e e^{\prime}}+V_{e e^{\prime}}G_{e^{\prime}}(R)
\right]
+ \sum_{g^{\prime}} V^{\text{int}}_{e g^{\prime}} G_{g^{\prime}}(R) \nonumber  \\
 =  [E-(n-1)\hbar \omega ] G_{e}(R). 
\end{eqnarray}
Here 
\begin{eqnarray}
T^G_{a^{\prime} a} & = & -\frac{\hbar^{2}}{2 \mu }\langle \Phi_{a^{\prime}}|
\frac{\partial ^{2}}{\partial R^{2}} G_{a}(R)|\Phi_{a} \rangle,
\label{cw-7} \\
V_{a^{\prime} a} & = & \langle \Phi _{a^{\prime}}|\left[ 
\hat{H}_{\text{rot}} + \hat{H}_{\text{el}} + \hat{H}_{\text{fs}}\right]
 | \Phi_{a} \rangle  ,
\label{cw-8}
\end{eqnarray}
($a=g$ or $e$) and
\begin{equation}
\label{cw-9}
V^{\text{int}}_{eg}= \langle \Phi_{e}, n-1|\hat{H}_{\text{int}}|\Phi_{g}, n \rangle 
= \sqrt{\frac{I}{2\epsilon_{0}c}}\;\langle \Phi_{e}|\bm{\epsilon}_{\lambda} \cdot \bm{d} 
|\Phi_{g}\rangle  
\end{equation}
where $I$ is the laser intensity and the molecular dipole operator $\bm{d}$ is the sum of 
the atomic dipole operators \mbox{$\bm{d}_{i}=-e\bm{r}_{i}$}. In obtaining (\ref{cw-9})
we have used \cite{Mies81}
\begin{equation}\label{cw-9a}
\langle \Phi_{e}|\frac{e}{m}\sum_{i}\hat{\bm{p}}_{i}|\Phi_{g} \rangle
=\frac{i}{\hbar} \langle \Phi_{e}|[\hat{H}_{\text{mol}},\bm{d}]|\Phi_{g}\rangle ,
\end{equation}
valid for the barycentric frame, and have invoked the dipole approximation 
$\exp (i \bm{k}\cdot \bm{r}_{i}) \approx 1$. This approximation has been used in 
other studies of photoassociation \cite{Portier06,Burke99} but does warrant some discussion.
The outer turning points of 
the $J=1$, $0_u^+$ vibrational states considered here lie in the range $(250-470)a_{0}$,
and $k =1/3258.17 \, a_0^{-1}$ for the  $2s\,{}^3S-2p\,{}^3P$ transition so that
the neglected next order term that would contribute is 
$ (kr_{1,2})^{2} \approx (kR/2)^{2} \sim 5 \times 10^{-3}$, comparable to other 
errors in our calculation. 

For photoassociation, the energy conservation relations are 
$E-n\hbar \omega =E^{\infty}_{g}+E_{K}$ and
$E-(n-1)\hbar \omega = E^{\infty}_{e}-B^{v}_{e}+ E_K + \hbar \Delta \omega $
where $E_{K}=\hbar^{2}k^{2}/2 \mu $ is the kinetic energy of the colliding atoms, 
$E^{\infty}_{g,e}$ are the energies of the asymptotically free atoms, and 
$\Delta \omega = \omega - \omega_{0}$ is the laser detuning. Here 
$\hbar \omega_0 = E^\infty_e - B^v_e - E^\infty_g$ is the separation of the 
unperturbed excited and ground state energies in the limit of zero kinetic energy.

\subsection {Basis states}

For two colliding atoms with orbital $\hat{\bm{L}}_{i}$, spin 
$\hat{\bm{S}}_{i}$ and total $\hat{\bm{j}}_{i}$ angular momenta, where 
$i ={1,2}$, several different basis representations can be constructed. 
Two possibilities are the $LS$ 
coupling scheme $\hat{\bm{L}}=\hat{\bm{L}}_{1}+\hat{\bm{L}}_{2}$, 
$\hat{\bm{S}}=\hat{\bm{S}}_{1}+\hat{\bm{S}}_{2}$ and 
$\hat{\bm{J}}=\hat{\bm{L}}+\hat{\bm{S}}+\hat{\bm{l}}$, and the $jj$ coupling 
scheme $\hat{\bm{j}}_{1}=\hat{\bm{L}}_{1}+\hat{\bm{S}}_{1}$,
$\hat{\bm{j}}_{2}=\hat{\bm{L}}_{2}+\hat{\bm{S}}_{2}$, $\hat{\bm{j}}=
\hat{\bm{j}}_{1}+\hat{\bm{j}}_{2}$ and $\hat{\bm{J}}=\hat{\bm{j}}+\hat{\bm{l}}$. 
The $LS$ coupling scheme diagonalizes $\hat{H}_{\text{el}}$ whereas the $jj$ coupling
scheme diagonalizes  $\hat{H}_{\text{fs}}$.
As the magnitude of the fine structure interaction is significantly larger
than the electronic interaction at the long ranges of the photoassociated molecules, 
we use the $jj$ coupled states $|\gamma j_{1}j_{2}jlJm_{J} \rangle $ (see Appendix A), 
where $\gamma$ represents other relevant quantum numbers 
and  $m_{J}$ labels the projections of the 
total angular momentum $\hat{\bm{J}}$ on to the space-fixed $Oz$ axis.
A further consideration is that the selection rules of the laser interaction refer to
the space-fixed reference frame and couple states of differing $J$ and $m_{J}$, 
whereas the molecular interactions are more naturally described in the molecular
reference frame in terms of $\Omega_{j}$, the projection of $j$ along the 
inter-molecular axis $OZ$. Hence we choose the hybrid Hund case (c) molecular basis   
\begin{equation}
\label{cw-10}
|\Phi_{a}(R,q) \rangle \equiv
 |\gamma_{1}\gamma_{2}j_{1}j_{2}j\Omega_{j}w,Jm_{J}\rangle 
\end{equation}
where $w$ is the symmetry under inversion of the electronic wavefunction through the 
centre of charge (see \text{Appendix A}).

The matrix elements of the various contributions to the Hamiltonian in this basis are
derived in Appendix B.

\section{Perturbative Treatment}

\subsection{F-operator technique}

For low laser intensities $\hat{H}_{\text{int}}$ can be treated as a perturbation.
The unperturbed Hamiltonian is then $\hat{H}_{0}=\hat{H}_{\text{mol}}
+\hat{H}_{\text{rad}}$ and has eigenstates of the form 
$R^{-1}G_i^0 (R) |\psi^{0}_{i}\rangle \otimes |n,\omega, \bm{\epsilon}_{\lambda} \rangle $ 
where the unperturbed radial functions $G^{0}_{i}(R)$ 
satisfy (\ref{cw-6}) with $V^{\text{int}}_{eg}$ omitted and $E$ replaced by the total 
unperturbed energy $E_{0}$. 

To calculate the second-order laser-induced energy level shifts $\delta E^{(2)}_{e}$ 
of the excited 
bound states $|e\rangle $ we note that \cite{Venturi03} have shown that the 
adiabatic approximation has little effect on the bound levels, making it 
possible to consider each excited state as an isolated state. 
The adiabatic states are formed by diagonalizing $\hat{H}_\text{mol}$ without 
the radial kinetic term and can be written as
\begin{equation}
\label{cw-20}
|\psi^{0}_{e} \rangle = \sum_{a}C_{ea}(R)|\Phi_{a} \rangle 
\end{equation} 
where $\sum_a |C_{ea}(R)|^2 = 1$.

We employ the F-operator technique \cite{Dal55,Merz98} to evaluate 
$\delta E^{(2)}_{e}$. This approach has recently been used by 
Beams et al. \cite{Beams04,Beams06} 
to treat spin-dipole interactions as a perturbation and we adapt their formalism 
here to the laser-matter interaction $\hat{H}_{\text{int}}$. The shift is given by
\begin{eqnarray}
\label{cw-25}
\delta E^{(2)}_{e} & = & \langle e|\hat{H}_{\text{int}}\hat{F}|e\rangle-
\langle e|\hat{F}|e\rangle\langle e|\hat{H}_{\text{int}}|e \rangle   
\nonumber \\
 & = &\langle e|\hat{H}_{\text{int}}\hat{F}|e\rangle,
\end{eqnarray}
where the second term is zero as $\langle e|\hat{H}_{\text{int}}| e \rangle $ 
vanishes due to dipole selection rules.  The operator $\hat{F}$ satisfies
\begin{equation}
\label{cw-24}
[\hat{F},\hat{H}_{0}]|e\rangle=\hat{H}_{\text{int}}|e\rangle .
\end{equation}

We assume the states $\{|\Phi_{g^{\prime}},n \rangle \}$ and 
$|\psi^{0}_{e}, n-1 \rangle \equiv |\psi_e^0 \rangle \otimes |n-1,\omega,\bm{\epsilon}_{\lambda}\rangle $
form a basis and expand $\hat{F}|e\rangle $ in terms of them:
\begin{equation}
\label{cw-25b}
\hat{F}|e\rangle = R^{-1} \left[ \sum_{g^{\prime}}f_{g^{\prime}}(R) 
|\Phi_{g^{\prime}},n \rangle + f_{e^{\prime}}(R) |\psi_{e^{\prime}}^0, n-1 \rangle \right],
\end{equation}
so that (\ref{cw-24}) gives the coupled equations
\begin{equation}
\label{cw-26}
\left[\frac{\hbar^{2}}{2\mu}\frac{d^{2}}{dR^{2}} + [E_{0}-(n-1)\hbar\omega] 
 - V^\text{KC}_{e} - 0_u^+(R) \right] f_e(R) = 0 
\end{equation}
and
\begin{eqnarray}
\label{cw-27}
\left[\frac{\hbar^{2}}{2\mu}\frac{d^{2}}{dR^{2}} + (E_{0}-n\hbar\omega)\right]
f_{g^{\prime}}(R)   \nonumber \\
 -\sum_{g}\left[V_{g^{\prime} g}^{\text{rot}}
+V_{g^{\prime} g}^{\text{el}}\right]f_{g}(R)= G^0_e(R) V_{g^{\prime}}^{\text{int}}(R).
\end{eqnarray}
We have assumed no $R-$dependence of the basis states $|\Phi_{a}\rangle $ but
include the correction term  $V^\text{KC}_{e}$ for the excited state (see Appendix B). 
The matrix elements are
\begin{eqnarray}
\label{cw-28}
V^{\text{el}}_{g^{\prime} g} & = & \langle \Phi_{g^{\prime}}|\hat{H}_{\text{el}}
|\Phi_{g} \rangle, 
\nonumber  \\
V^{\text{rot}}_{g^{\prime} g} & = & \langle \Phi_{g^{\prime}}|\hat{H}_{\text{rot}}
|\Phi_{g} \rangle
\end{eqnarray}
and $0_u^+(R)$ is the adiabatic potential of the excited state:
\begin{equation}
\label{cw-28b}
0_u^+(R) = \langle \psi^{0}_{e}| [\hat{H}_{\text{rot}} + \hat{H}_{\text{el}}
+ \hat{H}_{\text{fs}}]|\psi^{0}_{e} \rangle .
\end{equation}
The interaction term is
\begin{equation}
\label{cw-28a}
V^{\text{int}}_{g}(R) = \langle \Phi_{g}, n|\hat{H}_{\text{int}} |\psi^0_e, n-1\rangle
\end{equation}
and $G^0_e(R)$ is the radial eigenfunction for the rovibrational level $v$ of the
excited state.

Substitution of the expansion (\ref{cw-25b}) into (\ref{cw-25}) yields
\begin{equation}
\label{cw-28c}
\delta E^{(2)}_{e}= \sum_{g}\int dR \; G^{0}_{e}(R) V^{\text{int}}_{g}(R) f_{g}(R) .
\end{equation}
Consequently we need only solve the inhomogeneous equations (\ref{cw-27}). 
These are solved using a DVR method \cite{Light03} with a cosine Fourier basis.
As the laser interaction region requires closely spaced 
grid points and the asymptotic region very few points, a scaled grid 
$R=\zeta(t)$ \cite{Beams04} was used where the mesh in $t$ is equispaced. 
A quartic scaling was chosen as it does not modify the boundary conditions.

\subsection{Application to metastable helium}

We consider photoassociation to an excited long-range bound level in
the $J=1$, $m_J=1$, $0_u^+$ adiabatic potential that asymptotes to a 
He(2$s\;^3S_1$) + He(2$p\,^3P_0$) diatom (see Fig. \ref{fig:PAdiag}). 
The adiabatic state can be expressed as a combination of
the four basis states $|j_2,j\rangle$ 
\begin{equation}
\label{cw-33}
|0,1\rangle \quad |1,1\rangle \quad |2,1\rangle \quad |2,3\rangle
\end{equation}
where we have suppressed the common quantum numbers $\gamma_{1}=(L_1,S_1)=(0,1)$,
$\gamma_{2}=(L_2,S_2)=(1,1)$, $j_1=1$,  $\Omega_j=0$, $J=m_J=1$ and $w=1$. 
Coriolis couplings and the off-diagonal kinetic terms (see Appendix B) have been ignored, 
as their effect has been shown to be small at long range \cite{Venturi03}. 

An analysis of $\hat{H}_{\text{int}}$ shows that only a small manifold of metastable 
basis states are coupled to the excited state. Using the reduced state notation of 
$|j\Omega_j,Jm_J\rangle$, the states coupled by $\sigma^-$ laser light are
\begin{eqnarray}
\label{cw-34}
& |00,22\rangle , \quad |20,22\rangle ,\quad |21,22\rangle , \quad |2{-1},22 \rangle ,
\nonumber  \\
& (|22,22 \rangle ,\quad |2{-2},22 \rangle )
\end{eqnarray}
whereas the states coupled by $\sigma^+$ laser light are
\begin{eqnarray}
\label{cw-35}
&|00,00\rangle ,\quad |00,20\rangle ,\quad |20,00\rangle ,\quad |20,20\rangle , \nonumber \\
&|21,10\rangle , \quad  |2{-1},10 \rangle ,\quad |21,20\rangle ,\quad |2{-1},20 \rangle , 
\nonumber \\ 
&\left( |20,10\rangle , \quad |22,20\rangle ,\quad |2{-2},20 \rangle \right) .
\end{eqnarray}
The bracketed states are coupled by the Coriolis terms of
$\hat{H}_{\text{rot}}$ to other metastable states and hence are indirectly
coupled to the excited state. 
The implicit quantum numbers for these metastable states are 
\mbox{$\gamma_{1}=(L_1,S_1)=(0,1)$}, 
\mbox{$\gamma_{2}=(L_2,S_2)=(0,1)$}, \mbox{$j_1=j_2=1$} and $w=0$.

The calculations require Born-Oppenheimer potentials for the ${}^{5}\Sigma^{+}_{g}$ 
and ${}^{1}\Sigma^{+}_{g}$ states of the two metastable He(2$s\,{}^{3}$S$_{1}$)
atoms and the $J=1, 0^{+}_{u}$ potential of the excited He(2$s\,{}^{3}$S$_{1}$) +
He(2$p\,{}^{3}$P$_{0}$) system. For the quintet potential ${}^{5}\Sigma^{+}_{g}$ 
we use that given by Przybytek and Jeziorski \cite{Przyb05}, which
includes adiabatic, relativistic and QED corrections.
For the singlet potential ${}^{1}\Sigma^{+}_{g}$ 
we use \cite{Beams04} a potential constructed from the short-range
M\"uller \textit{et al}~\cite{Muller91} potential  exponentially
damped onto the quintet potential at long range. The excited state potentials are 
constructed from the twelve Born-Oppenheimer dispersion potentials \cite{Venturi03} 
$f_{3\Lambda}(R/\lambdabar) C_{3\Lambda}/R^3+ C_{6\Lambda}/R^6 + C_{8\Lambda}^\pm / R^8$, 
where $f_{3\Lambda}$ is an 
$R$- and $\Lambda$-dependent retardation correction~\cite{Meath68}, $\lambdabar=3258.17 \, a_0$, 
where $\lambdabar=\lambda/(2\pi)$ and $\lambda$ is the  $2s\,{}^3S-2p\,{}^3P$ transition 
wavelength. In particular, the excited state $|0^{+}_{u}\rangle $ 
is a linear combination of the ${}^{5}\Sigma^{+}_{u}$, ${}^{5}\Pi_{u}$,  ${}^{3}\Pi_{u}$, 
and  ${}^{1}\Sigma^{+}_{u}$ states. The $C_{3\Sigma}$ coefficient is 
$\pm 2 C_3$ and $C_{3\Pi}$ is $\pm C_3$, where $C_3 =6.41022 \, E_h a_0^3$ and is
proportional to the square of the $2s\,{}^3S-2p\,{}^3P$  transition dipole matrix element.
For the van der Waals coefficients we use $C_{6\Sigma}=2620.76 \,E_h a_0^6$ 
and $C_{6\Pi}=1846.60 \,E_h a_0^6$. The $C^\pm_{8\Lambda}$ terms are \cite{Marinescu} 
$C^+_{8\Sigma}=151383 \,E_h a_0^8$, $C^-_{8\Sigma}=297215.9 \,E_h a_0^8$, 
$C^+_{8\Pi}=97244.75 \,E_h a_0^8$ and $C^-_{8\Pi}=162763.8 \,E_h a_0^8$. Here  the 
superscript indicates the sign of $(-1)^{S+w}$ where $S$ is the spin of the state and
$w=0(1)$ for gerade (ungerade) symmetry. The $C_3/R^3$ term is the dominant 
contribution to the dispersion potential for the purely-long-range states. 
Also required is the $2s \rightarrow 2p$ atomic dipole moment 
$d_{\text{at}}^{\text{sp}}= 2.14583 \times 10^{-29}$ C.m and the fine structure splittings. 
For He the $2p \, {}^3P_0$ and $2p \, {}^3P_1$ levels lie $31.9081$ GHz and $2.2912$ GHz above 
the $2p \, {}^3P_2$ level, respectively.

The results for the three lowest vibrational bound levels and for laser polarizations 
$\sigma^{\pm}$ are given in table \ref{tab:perturb}. In order to compare with the results of Portier
et al. \cite{Portier06}, we have used $\Delta \omega =0$, corresponding to a zero kinetic energy.
There is good general agreement between the two sets of results with
the small differences probably due to the potentials used.

\begin{table}
\caption{\label{tab:perturb} Perturbative line shifts per laser intensity 
in units of MHz/(W.cm$^{-2}$) for the long range bound states of the 
\mbox{He($2s\,^3S_1$) + He($2p\,^{3}P_{1}$), $0_{u}^{+}$} configuration. Results
are given for calculations without and with the correction $V^\text{KC}_e$.} 
\begin{ruledtabular}
\begin{tabular}{lcccc}
 Level & Polarization & No $V^\text{KC}_e$ & With $V^\text{KC}_e$ & Ref \cite{Portier06}  \\
\hline
$v=0$ & $\sigma^{-}$ & -6.439 & -6.507 & -6.37  \\ 
 & $\sigma^{+}$ & -7.724 & -7.784 & -7.36   \\ 
$v=1$ & $\sigma^{-}$ & -11.662 & -11.748 & -11.70  \\ 
 & $\sigma^{+}$ & -10.205 & -10.270 & -10.25  \\
$v=2$ & $\sigma^{-}$ & -29.442 & -29.692 & -29.57  \\
 & $\sigma^{+}$ & -24.675 & -24.877 & -24.11 \\
\end{tabular}
\end{ruledtabular}
\end{table}

\section{Close-coupled calculation of photoassociation profile}

\subsection{Hamiltonian matrix}

The close-coupled calculation is non-perturbative in that
the differential equations (\ref{cw-6}) are solved without assuming the 
laser interaction is weak. In this approach the scattering matrix elements 
are calculated and the PA profile obtained for various laser detunings, 
intensities and temperatures. 

Although the method presented here is quite general,
for ease of visualization we explicitly formulate it for photoassociation
in metastable helium of the subset of states coupled by $\sigma^-$ polarized light in (\ref{cw-34}). 
Specifically these
states $|\alpha \rangle $ are
\begin{eqnarray}
\label{cw-40}
|1\rangle &=& |j=0,\Omega_j=0,J=2,m_J=2\rangle \otimes |n,\omega, \bm{\epsilon}_{\lambda} \rangle , \nonumber \\
|2\rangle &=& |j=2,\Omega_j= - 2,J=2,m_J=2\rangle \otimes |n,\omega, \bm{\epsilon}_{\lambda} \rangle , \nonumber \\
|3\rangle &=& |j=2,\Omega_j= - 1,J=2,m_J=2\rangle \otimes |n,\omega, \bm{\epsilon}_{\lambda} \rangle , \nonumber \\
|4\rangle &=& |j=2,\Omega_j=\phantom{-}0,J=2,m_J=2\rangle \otimes |n,\omega, \bm{\epsilon}_{\lambda} \rangle , \nonumber \\
|5\rangle &=& |j=2,\Omega_j=\phantom{-}1,J=2,m_J=2\rangle \otimes |n,\omega, \bm{\epsilon}_{\lambda} \rangle , \nonumber \\
|6\rangle &=& |j=2,\Omega_j=\phantom{-}2,J=2,m_J=2\rangle \otimes |n,\omega, \bm{\epsilon}_{\lambda} \rangle , \nonumber \\
|7\rangle &=& |0_u^+,J=1,m_J=1\rangle \otimes |n-1,\omega, \bm{\epsilon}_{\lambda} \rangle  .
\end{eqnarray}
for which the close-coupled equations (\ref{cw-6}) reduce to
\begin{equation}
\label{cw-41}
\sum_{\alpha^{\prime}}\left[-\frac{\hbar^{2}}{2 \mu } \frac{d^{2}}{dR^{2}}\;
\delta_{\alpha \alpha^{\prime}} +W_{\alpha \alpha^{\prime}}(R)\right] 
G_{\alpha^{\prime}}(R) =0
\end{equation} 
where the potential matrix $\bm{W}$ is
\begin{widetext}
\begin{equation}
\label{cw-42}
\bm{W} = \left[\begin{array}{ccccccc}
\! V_{1}^0 - E_K\! & 0 & 0 & 0 & 0 & 0 & \hbar\Omega_{1}^0 \\
0 & \! V_{5}^{-2} - E_K\! & L^{12} & 0 & 0 & 0 & 0 \\
0 & L^{12} &\! V_{5}^{-1}- E_K\! & L^{01} & 0 & 0 & \hbar\Omega_{5}^{-1} \\
0 & 0 & L^{01} &\! V_{5}^0- E_K\! & L^{01} & 0 & \hbar\Omega_{5}^0 \\
0 & 0 & 0 & L^{01} &\! V_{5}^1- E_K\! & L^{12} & \hbar\Omega_{5}^1 \\
0 & 0 & 0 & 0 & L^{12} & \! V_{5}^2- E_K\! & 0 \\
\hbar\Omega_{1}^0 & 0 & \hbar\Omega_{5}^{-1} & \hbar\Omega_{5}^{0} & \hbar\Omega_{5}^{1} 
& 0 & V_{e}+V^{\text{KC}}_{e}-\Delta E
\end{array}\right].
\end{equation}
\end{widetext}
Here 
\begin{equation}
\label{cw-43}
V^{\Omega_j}_{2j+1}= {}^{2j+1}\Sigma_g^+ (R) + \langle j\Omega_j | \hat{H}_\text{rot} | j\Omega_j \rangle,
\end{equation}
is the effective ground state potential, comprising the Born-Oppenheimer potential for the 
${}^{2j+1}\Sigma_g^+ $ state and the diagonal rotational couplings, $V_e = 0^{+}_{u}(R) -i\Gamma /2$ is 
the adiabatic potential of the excited state, and $\Delta E = \hbar \Delta \omega + E_K - B^v_e$. 
The off-diagonal terms are the Coriolis couplings
\begin{equation}
\label{cw-43a}
L^{\Omega_j,\Omega_j^\prime} = \langle j=2,\Omega_j^\prime | \hat{H}_\text{rot} | j=2,\Omega_j \rangle 
\end{equation}
and the radiative couplings $\hbar \Omega_{2j+1}^{\Omega_j} = V^{\text{int}}_{g}(R)$ expressed in 
terms of the Rabi frequencies $\Omega_{2j+1}^{\Omega_j}$. 

The decay of the excited state through spontaneous emission is represented by
the molecular width $\Gamma = \Gamma(R/\lambdabar, \Gamma_{\text{at}})$
where the $R$-dependence arises from the retarded dipole interaction expressed
in the Hund's case (a) states \cite{Meath68} and its subsequent diagonalization as part of
the formation of the adiabatic potential. For our \textit{ungerade} system, 
$\Gamma \approx 2 \Gamma_{\text{at}}$ for the interaction region and 
$\Gamma \approx \Gamma_{\text{at}}$ for the asymptotic region $R \gg \lambdabar $. 
The spontaneous emission atomic line width for the triplet helium 2s-2p
transition is $\Gamma_\text{at} = 1.626$ MHz. 
Note that the asymptotic energies $E^{\infty}_{g,e}$ cancel in (\ref{cw-42}).

The definition of scattering matrix elements requires asymptotically free states. However the
asymptotic form $\bm{W}_\infty $ of $\bm{W}$ is not diagonal as the Rabi couplings
do not vanish at large $R$ and the separated atoms
remained coupled by the laser interaction. Two options are available to determine the $S$-matrix 
elements: (i) transformation from the basis states
$|\alpha \rangle $ to the dressed-state basis states \cite{Mies81} 
$|\beta \rangle = \sum_{\alpha}U_{\beta \alpha }|\alpha \rangle $ in which 
\mbox{$\bm{W}_{\infty}^{\text{D}}=\bm{U}^{-1}\bm{W}_{\infty}\bm{U}$} is diagonal or, 
(ii) introduction of a modified radiative coupling for $R$ greater than some large value $R_{z}$:  
\begin{equation}
\label{cw-44a}
\tilde{V}_{eg}^{\text{int}}(R)= \left\{ \begin{array}{ll}
	V_{eg}^{\text{int}}(R), & \textrm{for}\;R \leq R_{z} \\
	V_{eg}^{\text{int}}(R) \exp{[-\rho (R-R_z)^{2}]}, &  \textrm{for}\;R > R_{z}
        \end{array} \right. .
\end{equation}
This coupling vanishes asymptotically and simulates the experimental conditions of the laser 
being switched off before and after the experiment.

\subsection{Dressed States}
 
Napolitano \cite{Napolitano98} has considered the case of six dressed states with laser couplings 
restricted to two pairs of states
and has obtained analytical results valid for large red 
detunings $\Delta E \gg \Gamma $ and $\Delta E \gg \Omega^{\Omega_{j}}_{5}$. As we
wish to avoid such assumptions and also develop a procedure valid for a larger number of coupled
states, we use direct numerical diagonalization.

In terms of dressed states the expansion of the state vector becomes
\begin{equation}
\label{cw-45}
|\Psi \rangle = R^{-1} \sum_{\alpha} G_{\alpha}(R) |\alpha \rangle = 
R^{-1} \sum_{\beta }\tilde{G}_{\beta}(R) |\beta \rangle  
\end{equation}
where the dressed-state radial amplitudes satisfy  
\begin{equation}
\label{cw-46}
\sum_{\beta^{\prime}}\left[-\frac{\hbar^{2}}{2 \mu } \frac{d^{2}}{dR^{2}}\;
\delta_{\beta \beta^{\prime}} +W^{\text{D}}_{\beta \beta^{\prime}}(R)\right] 
\tilde{G}_{\beta^{\prime}}(R) =0
\end{equation} 
and $\bm{W}^{\text{D}}=\bm{U}^{-1}\bm{W}\bm{U}$. Asymptotically the dressed states
decouple and satisfy
\begin{equation}
\label{cw-47}
\left[ \frac{d^{2}}{dR^{2}} + \frac{2 \mu E_{\beta }}{\hbar^{2}}\right] 
\tilde{G}_{\beta}(R) =0
\end{equation}
where the energies $E_{\beta } \equiv -W^{\text{D}}_{\beta \beta }(R= \infty)$ 
can be complex. Defining $k_{\beta } =\sqrt{2 \mu E_{\beta }}/\hbar =
k^{r}_{\beta} + i k^{i}_{\beta }$, the asymptotic solutions of (\ref{cw-47})
have the form
\begin{equation}
\label{cw-48}
\tilde{G}_{\beta} (R) \sim c_{1}e^{-k_{\beta}^{i}R} e^{ik_{\beta}^{r}R}
+ c_{2}e^{k_{\beta}^{i}R} e^{-ik_{\beta}^{r}R}.
\end{equation}
The open-channel and closed-channel manifolds are then identified 
respectively as those states that persist or vanish as $R \rightarrow \infty $.
Therefore we require $k_{\beta }^{i}$ to be zero for the open channels.

After diagonalisation we find that two of the energies $E_{\beta }$
are nondegenerate and their associated dressed states include undressed excited
state contributions. The third value of $E_{\beta }$ is purely real, five-fold
degenerate and mixes only the undressed open channels for the ground states.
As the diagonalization is numerical, the particular combination of ground
states in this degenerate subspace is dependant upon the numerical procedure
used and the combinations will \emph{not} vary smoothly as the detuning is
changed.  However, the \mbox{$S$-matrix} elements that involve only the states
with non-degenerate energies will be guaranteed to vary smoothly.

For the orthogonal polarization $\sigma^+$ we briefly note that the above
analysis applied to the 11 metastable states listed in \eqref{cw-35} exhibits very similar
behaviour. Three unique values for $E_\beta$ are observed, two that are
non-degenerate and one that is 10-fold degenerate. No differences in the
analysis or numerical implementation apply other than to include a larger number
of states in the $\sigma^+$ system.

\subsection{Modified Radiation Coupling}

Because the dressed states introduce complex asymptotic energies, the simplicity
offered by using a modified radiative coupling becomes more attractive. To implement
the modified coupling, we use the undressed form of the state vector
\eqref{cw-45} to obtain
\begin{equation}
\label{cw-49}
\sum_{\alpha^{\prime}}\left[-\frac{\hbar^{2}}{2 \mu } \frac{d^{2}}{dR^{2}}\;
\delta_{\alpha \alpha^{\prime}} +W^{z}_{\alpha \alpha^{\prime}}(R)\right] 
\tilde{G}_{\alpha^{\prime}}(R) =0
\end{equation}
where $W^{z}_{\alpha \alpha^{\prime}}$ is identical to $W_{\alpha
\alpha^{\prime}}$ except that $V^\text{int}_{eg}$ is replaced by
$\tilde{V}^\text{int}_{eg}$. This system
consists of six open channels, all with identical values
of $k_\alpha = \sqrt{2\mu E_K} / \hbar$, and one closed channel.

\subsection{Extraction of $S$-matrix elements}

The $S$-matrix is determined by matching the asymptotic solutions of (\ref{cw-46})
or (\ref{cw-49}) to the combination \cite{Mies80}
\begin{equation}
\label{cw-60}
\bm{\tilde{G}}=\bm{H}_{-}^{0}\bm{A}+\bm{H}_{+}^{0}\bm{B} = 
(\bm{H}^{0}_{-} - \bm{H}^{0}_{+}\bm{S})\bm{A}
\end{equation}
where $\tilde{G}_{\gamma \gamma^{\prime}}$ is the matrix of solutions formed from
$\tilde{G}_{\gamma}(R)$ with the second subscript 
$\gamma^{\prime}$ labelling the linearly independent solutions generated by different choices 
of boundary conditions. We have introduced the notation $\gamma = \beta $ for the 
dressed states approach and $\gamma = \alpha $ for the modified coupling approach.
The diagonal matrices $(H^{0}_{\pm})_{\gamma \gamma^{\prime}}=
\delta_{\gamma \gamma^{\prime}}h^{\pm}_{\gamma}$ for the open channel scattering states 
have the asymptotic form of outward and inward travelling waves:  
\begin{equation}
\label{cw-61}
h_{\gamma}^{\pm}  \underset{R\rightarrow\infty}{\sim}(2|k_{\gamma}|)^{-\frac{1}{2}}
e^{\pm ik_{\gamma}R}
\end{equation}
whereas for the closed channels the asymptotic forms are
\begin{equation}
\label{cw-62}  
h_{\gamma}^{\pm}  \underset{R\rightarrow\infty}{\sim} (2|k_{\gamma}|)^{-\frac{1}{2}}
e^{\mp |k_{\gamma}|R}.
\end{equation}

The matrices $\bm{A}$, $\bm{B}$, and $\bm{S}=-\bm{B}\bm{A}^{-1}$ have the structure 
\begin{equation}
\label{cw-63}
\bm{A} = \left[ \begin{array}{cc} 
	\bm{A}_{oo} & \bm{A}_{oc} \\
	\bm{A}_{co} & \bm{A}_{cc} \end{array} \right]
\end{equation}
where the labels $o$ and $c$ refer respectively to open and closed channels. 
We require only the open-open contributions $\bm{S}_{oo}$ which, as we can 
enforce $\bm{A}_{co}=0$ by careful choice of boundary
conditions, are given by $\bm{S}_{oo}=-\bm{B}_{oo} \bm{A}^{-1}_{oo}$.

As $E_{\beta} = E_{\beta}^r + i E_{\beta}^i$ and $k_{\beta}$ can be complex for the dressed states, 
the asymptotic solutions do not have the usual plane 
wave oscillatory form and care must be taken during the matching process. In the present case
$E_\beta$ is complex for only two of the dressed states. One is clearly a closed channel as it has 
$E_\beta^r < 0$. The other has $E_\beta^r>0$ and a relatively small value of $E_{\beta}^i$.
Without the imaginary component, this channel would have a purely real value of $k_\beta$
and be considered an open channel. However, as $k_{\beta}$ is complex, the state has the 
asymptotic form (\ref{cw-48}), consisting of exponentially increasing and decreasing solutions. 
To treat the solution rigorously, we must enforce finiteness by discarding the exponentially increasing 
solution and integrate out to a sufficiently large value of $R$ that the exponentially decreasing solution has 
completely dampened, indicating that the channel is closed. The $S$-matrix can then be created from 
the remaining open channels that have purely real $E_\beta$ components.
The difficulty with such an approach is that $k_\beta^r \gg k_\beta^i$,
requiring the integration range to be very large, imposing a heavy computational
burden. Note that $k_\beta^r$ and $k_\beta^i$ will also vary with the laser
detuning and laser intensity, further complicating the problem.

An alternative approach is to consider the problematic channel as a pseudo-open channel. By 
redefining the S-matrix, such that the asymptotic functions are matched at a
finite distance $R=R_\text{max}$ instead of $R \rightarrow \infty$, we can ensure
the pseudo-open channel has a finite inward and outward flux. This is performed
by treating the terms $e^{\mp k^i_\beta R}$ as approximately constant in the local
region of $R_\text{max}$ and matching to the oscillatory behavior of $e^{\pm i k^r_\beta R}$.
This type of matching obviously has a dependence upon the matching point but
we expect it to not vary the shape of the profile significantly if $R_\text{max}$ is chosen outside the 
interaction region. 
The resultant $S$-matrix element $S_{\beta^{\prime}\beta}$ then gives the probability that the system 
with unit flux in an incoming channel $|\beta \rangle $ at $R_\text{max}$ makes a
transition to an outgoing channel $|\beta^{\prime}\rangle $ at $R_\text{max}$.
We note that this choice of $R_\text{max}$ bears some similarity to the choice
of $R_z$ for the modified coupling method.

\subsection{Detection of the Resonance}

The photoassociation resonance can be studied by analysing the loss from the excited state due 
to spontaneous emission so that the photoassociation lineshape is due to the emitted photons 
and therefore proportional to the loss of
unitarity of the $S$-matrix. For atoms colliding in the entrance channel
$|\gamma \rangle $, the loss rate is
$\mathcal{L}_{\gamma}=\langle v_{\gamma}\sigma^{\text{photon}}_{\gamma}\rangle $ where the 
cross section for photon emission is
\begin{equation}
\label{cw-51}
\sigma^{\text{photon}}_{\gamma}  = \frac{\pi }{k_{\gamma}^{2}}
\left(1-\sum_{\gamma^{\prime}}|S_{\gamma^{\prime} \gamma}|^{2}\right)
\end{equation}
and $\langle \cdots \rangle $ denotes a thermal average over a
distribution of the  asymptotic relative velocities
$v_{\gamma }=\hbar k_{\gamma}/\mu $ of the two colliding atoms. For temperatures of order 
1 $\mu $K this thermal averaging can be ignored.
The energy dependence of $\sigma^{\text{photon}}_{\gamma}$ is that of a peak superimposed upon a 
slowly varying background and can be well fitted in the region of the peak by a
Fano profile \cite{Fano61} of the form
\begin{equation}
\label{cw-52}
\sigma^{\text{photon}}_{\gamma}=A_\text{bg}(\epsilon) - A_\text{res} \frac{(\epsilon + q)^2}{1 + \epsilon^2}
\end{equation}
where $A_\text{bg}(\epsilon)$ describes a linear background, $A_\text{res}$ is a 
constant, $\epsilon = (E - E_\text{res}) / (\Gamma_\text{res} / 2)$ is a
normalized energy, and $E_\text{res}$ and $\Gamma_\text{res}$ are the position and full width of the
resonance. The Fano parameter $q$ is a measure of the ratio of the direct (background) to 
resonant scattering. As we often deal with Lorentzian-like behaviour that occurs
in the limit $q \rightarrow \infty$ and $A_\text{res} \rightarrow 0$, it is
simpler numerically to instead match to the form
\begin{equation}
\label{cw-52b}
\sigma^{\text{photon}}_{\gamma}=A_\text{bg}(\epsilon) - A^\prime_\text{res} \frac{(1 + p \epsilon)^2}{1 + \epsilon^2}
\end{equation}
where $p = 1/q$ and $A^\prime_\text{res} = q A_\text{res}$. For the present calculations we find
$p \alt 10^{-2}$ except for the dressed state profiles at high intensity where $p \sim 0.5$.

\subsection{Numerical issues}

To solve the dressed problem, the undressed asymptotic potential matrix $\bm{W}_\infty$ 
is diagonalized by a Hermitian 
eigendecomposition. The renormalized Numerov \cite{Johnson78} boundary value method 
is then used to integrate either the
dressed equations \eqref{cw-46} or the modified coupling equations
\eqref{cw-49}. The outer boundary used is either $R_\text{max}$ for the dressed states 
or, for the modified coupling, a value greater than $R_z$ such that the radiation
coupling has been completely turned off. To
ensure the asymptotic solutions are uncoupled, we choose $R_\text{max}$ and $R_z$
to be greater than $10^{4}$ $a_0$. Typically we use $R_{z} =2 \times 10^{5} a_{0}$
and $\rho \sim O(5 k_{\gamma}^{-1})=4 \times 10^{-6} a_{0}^{-1}$.  
The linearly independent solutions $\tilde{G}_{\gamma \gamma^\prime}$, 
$\gamma^\prime = 1,...,\mathcal{N}-1$, are generated by choosing 
$\mathcal{N} - 1$ linearly independent boundary conditions.
 
A kinetic energy of $10^{-11}$ $E_h$ (2.1 $\mu$K) is chosen, which places the
system just above temperatures for which recent experiments have 
reported Bose-Einstein condensation
and which does not introduce a prohibitively large outer boundary for
the numerical integration. 
Note that quantitatively, at an
intensity of 0.7 W/cm$^2$, the matrix elements in $\bm{W_\infty}$ are $O(10^{-11})$ and 
$O(10^{-7})$ for the diagonal elements $E_K$ and $\Delta E$ respectively and $O(10^{-9})$ for 
the off diagonal Rabi couplings $\Omega_{2j+1}^{\Omega_j}$.

\subsection{Results}

We first consider the results for low to moderate laser intensities as they
exhibit behavior similar to regular spectroscopic profiles. As the laser
intensity is increased, unusual aspects of the dressed and modified coupling
become apparent and we shall discuss these separately.

The central positions and broadenings of the dressed PA profiles were determined 
from the fits to the Fano profiles (\ref{cw-52}) for the cross section 
$ \sigma^{\text{photon}}_{\gamma}  $
as a function of laser energy. 
For nondegenerate channels $|\beta \rangle $, such as the pseudo-open channel 
in the present investigation, the variation of $\sigma^\text{photon}_\beta$ with laser 
energy is smooth. However, as previously mentioned, the numerical diagonalization process 
arbitrarily selects the degenerate dressed states $\{|\beta_{d} \rangle \}$ in the 
subspace $\mathcal{E}_{d}$ that they span. 
This means that $\sigma^\text{photon}_\beta$ for any $\beta \in
\beta_d$ will not vary smoothly with laser energy. 

\begin{figure}
\includegraphics[width=0.95\linewidth]{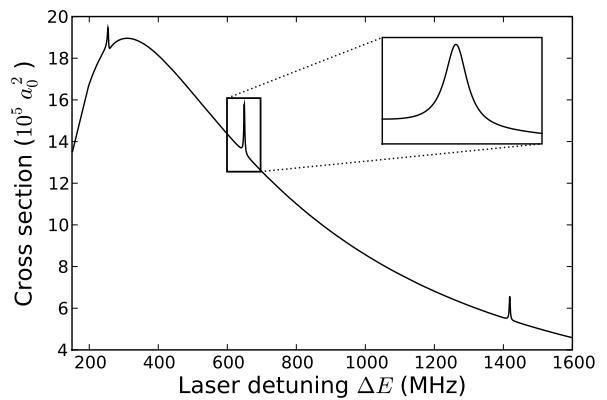} 
\caption{\label{fig:lowintd}
PA cross section profile calculated using dressed states. Results shown are for $\sigma^-$ polarization 
and a low intensity of 64 mW/cm$^2$.}
\end{figure}

Fortunately, for these degenerate dressed channels we are permitted to analyse various combinations of 
$\sigma^\text{photon}_\beta$ as the physical behaviour of the system 
cannot depend on the choice of basis
in $\mathcal{E}_{d}$. The simplest choice is an average of all channels, i.e.
\begin{equation}
\label{cw-70}
\sigma^\text{photon}_\text{dressed} = \frac{1}{1+n_{d}} \sum_{\beta}
\sigma^\text{photon}_\beta 
\end{equation}
where $n_{d}$ is the number of degenerate states.  Results for the case of 
$\sigma^{-}$ polarization and a low intensity of 64 mW/cm$^{2}$ are shown in 
Fig. \ref{fig:lowintd}.
 
\begin{figure}
\includegraphics[width=0.95\linewidth]{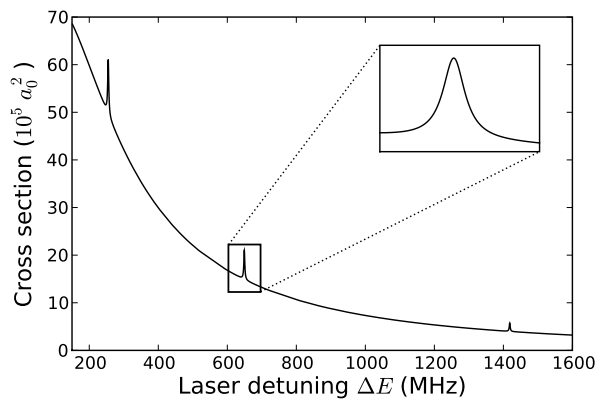} 
\caption{\label{fig:lowintm}
PA cross section profile calculated using the modified coupling. Results shown are for $\sigma^-$ polarization 
and a low intensity of 64 mW/cm$^2$.}
\end{figure}

For the modified coupling method, the cross section $\sigma^{\text{photon}}_{\alpha}$ for the undressed 
channel $|\alpha \rangle $ is more straight forward to analyse in terms of experimental conditions. 
We choose to form a similar quantity 
\begin{equation} 
\label{cw-71} 
\sigma^\text{photon}_\text{modified} = \frac{1}{n_o} \sum_\alpha \sigma^\text{photon}_\alpha
\end{equation}
where $n_o$ is the number of open channels, and again fit the resonances to Fano
profiles. In this way we can compare the profiles with those of the
dressed state calculation. Results for $\sigma^{-}$ polarization and an intensity of 
64 mW/cm$^{2}$ are shown in Fig. \ref{fig:lowintm}.

The two profiles shown in Figs \ref{fig:lowintd} and \ref{fig:lowintm} are very similar apart from the 
suppression of the background at small detunings for the dressed state profile.
Importantly, the resonance parameters $E_{\text{res}}$ and $\Gamma_{\text{res}}$ obtained
from fits to the two profiles are identical. The background loss behavior present in the 
profiles is due to the dominance of the off-diagonal Rabi couplings over the diagonal 
terms $\Delta E$ and $E_{K}$ of the potential matrix for the asymptotic region. 
This situation arises because of the shallow nature of the excited state potential and 
the ultracold temperature used. 
If the well is artificially deepened, or much larger kinetic
energies are used, then this background completely disappears.

\begin{figure}
\includegraphics[width=0.95\linewidth]{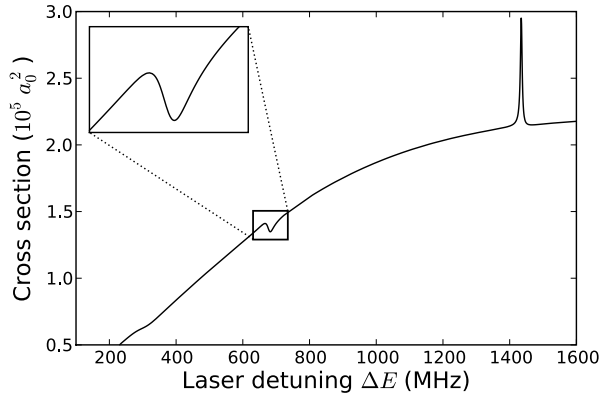}
\caption{\label{fig:dressedhighint}
PA cross section profile calculated using dressed states. Results shown are for $\sigma^-$ polarization 
and a high intensity of 2.6 W/cm$^2$.}
\end{figure}

\begin{figure}
\includegraphics[width=0.95\linewidth]{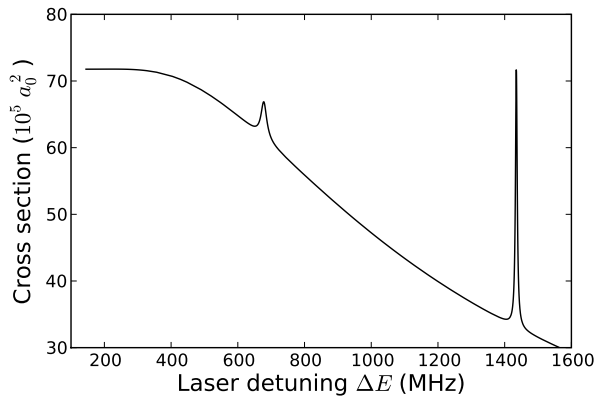}
\caption{\label{fig:rdephighint}
PA cross section profile calculated using the modified coupling. Results shown
are
for $\sigma^-$ polarization and a high intensity of 2.6 W/cm$^2$.}
\end{figure}

The PA profiles for high laser intensity show some unusual behavior. Results for the dressed state 
and modified coupling profiles at an intensity 
of 2.6 W/cm$^{2}$ are shown in Figs \ref{fig:dressedhighint} and \ref{fig:rdephighint} respectively.
These spectra exhibit two features not present in previously calculated PA profiles; strong
interference between the resonance and background contributions and a reduction in the
overall magnitude of the dressed profile. The interference feature is apparent in the
magnitude of the corresponding Fano parameter $p =1/q \sim 0.5$, indicating that 
these resonances exhibit a
severe departure from Lorentzian-like behavior.

\begin{figure}
\includegraphics[width=0.95\linewidth]{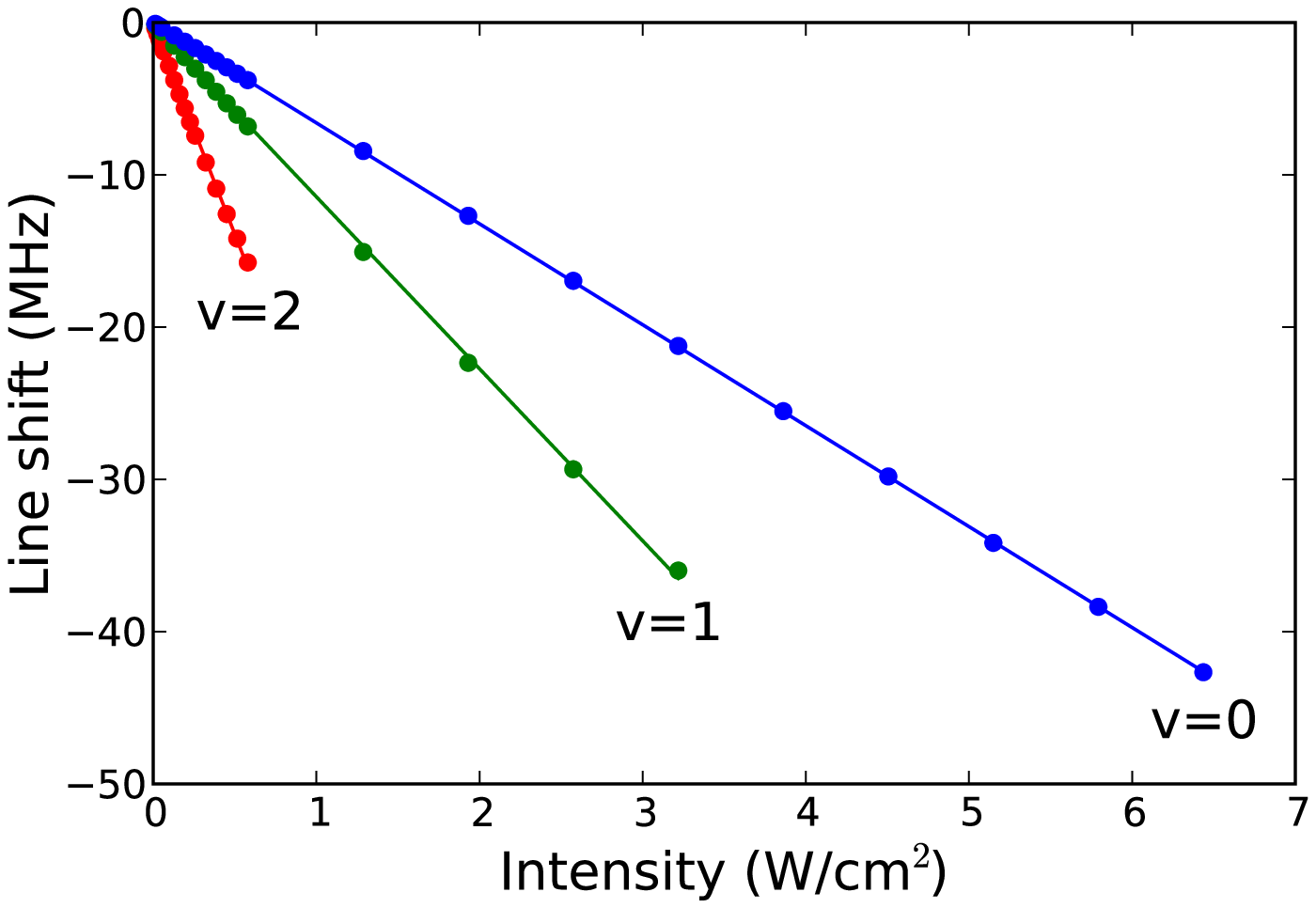}
\caption{(Color online) Dependence upon laser intensity of the line shift of the $v=0,1,2$ levels for
$\sigma^-$ polarization coupled to the $0_u^+$ state.}
\label{fig:lineshifts}
\end{figure}

\begin{figure}
\includegraphics[width=0.95\linewidth]{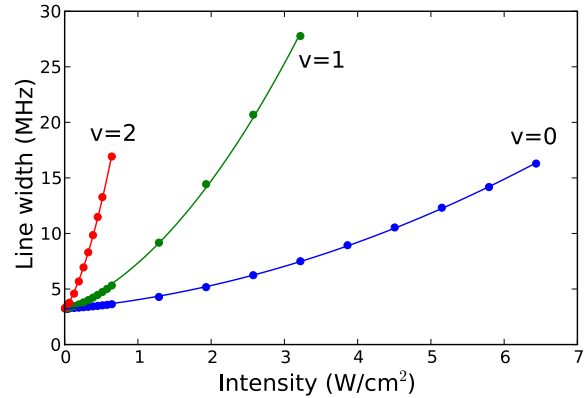}
\caption{(Color online) Dependence upon laser intensity of the line width of the $v=0,1,2$ levels for
$\sigma^-$ polarization coupled to the $0_u^+$ state.}
\label{fig:linewidths}
\end{figure}

The dependence of the resonance position and width upon laser intensity is shown in
Figs \ref{fig:lineshifts} and \ref{fig:linewidths} respectively for the case of $\sigma^{-}$ 
polarization. Similar behavior is obtained for $\sigma^{+}$ polarization. The intensity 
dependence of the line shift and width is very close to linear and quadratic respectively
for $v=0,1$ but departures from these dependencies are evident for $v=2$. 
This can be seen from the fit parameters given in Table \ref{tab:fits} which shows that the 
coefficients $s_{2}$ of the quadratic correction and $w_{3}$ of the cubic correction to 
the intensity dependence of the shift and width respectively for $v=0,1$ are very small. 

\begin{table*}
\caption{\label{tab:fits}Parameters for a quadratic fit $s_{1}I + s_{2}I^{2}$ and cubic fit 
$w_{0}+w_{1}I+w_{2}I^{2}+w_{3}I^{3}$ to the dependence of the line shifts and the line full
widths respectively (in MHz) upon laser intensity $I$ (W/cm$^{2}$). The value $I_{\text{max}}$
denotes the approximate maximum intensity up to which the resonance peak is clearly 
discernable and there is no overlap with neighboring peaks.}
\begin{ruledtabular}
\begin{tabular}{lcddddddd}
Level & Polarization & \multicolumn{2}{c} {Shift} & \multicolumn{4}{c}
{Width} & \multicolumn{1}{c} {$I_{\text{max}}$} \\
& & \multicolumn{1}{c}{$s_{1}$} & \multicolumn{1}{c} {$s_{2}$} & \multicolumn{1}{c} {$w_{0}$} &
\multicolumn{1}{c} {$w_{1}$} &   \multicolumn{1}{c} {$w_{2}$} & \multicolumn{1}{c} {$w_{3}$} & \\
\hline
$v=0$ & $\sigma^{-}$ & -6.514 & -0.023  & 3.233 & 0.441 & 0.304 & -0.0093 &  7.0 \\
$v=0$ & $\sigma^{+}$ & -7.781 & -0.021  & 3.226 & 0.730 & 0.329 & -0.0091 &  7.0 \\
$v=1$ & $\sigma^{-}$ & -11.77 &  0.074  & 3.231 & 1.51  & 2.87 & -0.34  &  3.2 \\
$v=1$ & $\sigma^{+}$ & -10.30 & -0.018  & 3.219 & 1.47  & 1.70 & -0.10  &  3.2 \\
$v=2$ & $\sigma^{-}$ & -29.79 &  3.90  & 3.216 & 6.25  & 36.9 & -21.9  &  0.4 \\
$v=2$ & $\sigma^{+}$ & -24.95 &  1.13  & 3.197 & 5.44  & 18.2 &  -4.20  &  0.6
\end{tabular}
\end{ruledtabular}
\end{table*}


\section{Discussion and conclusions}

We have investigated the line shifts and widths for photoassociation of 
spin-polarized metastable helium to the three lowest rovibrational levels of the 
$J=1$, $0_u^+$ state asymptoting to 2$s\,{}^{3}$S$_{1}+$2$p\,{}^{3}$P$_{0}$ using two variants 
of a non-perturbative close-coupled treatment, one based upon dressed states and the 
other on a modified radiative coupling which vanishes asymptotically. We have also 
calculated the shifts using a second order perturbative treatment. 

The main physical interest is in the properties of the PA resonance profiles and our results for 
the shifts and widths of these profiles obtained using the two 
variants of the non-perturbative calculation agree to better than 2\%.
Both methods indicate that there is significant background loss for this metastable helium 
system, a feature not present in studies of PA in other systems, which is due to the 
shallow nature of the excited state potential.

The behavior of the PA profiles for high laser intensity is quite different for the dressed state
and modified coupling methods.
Physically, the behavior of the modified coupling profile appears to make sense; if the laser is intense
enough that most loss occurs outside the collision region, then the
particular resonance of the excited state will have little effect on the profile. 
Hence a form of saturation is observed. Note that this first occurs
in the higher vibrational levels as the intensity is increased. 
The behavior of the dressed states profile is very different in that the overall magnitude of the 
profile decreases at higher intensities and lower
detunings. It is difficult to relate this behavior to the collision
process without the appropriate association of the dressed state description with the experimental, 
undressed states. When the laser is switched on, the initial undressed states must be transformed to the
dressed states and in this process the undressed open channels acquire significant components of the 
pseudo-open and closed dressed channels and thereby suffer loss. In the limit of infinite intensity, 
the open dressed channels
become completely uncoupled from the pseudo-open and closed channels, and the loss from the dressed
states is entirely due to the activation of the laser. This suggests
that the general interpretation of quantities constructed from dressed S-matrix elements at large intensities 
requires a proper description of the activation of the laser in the formalism. 
Fortunately, this detail is not required to obtain the resonance parameters as both treatments result in 
nearly identical shift and width values. 

The results for the line shifts from the close-coupled and perturbative
calculations agree very closely at low laser intensities. The $s_{1}$ values in Table \ref{tab:fits}
differ by $\alt 0.5\%$ from our perturbative results (see Table \ref{tab:perturb}) calculated with the 
correction $V^{\text{KC}}_{e}$ included. The small differences can be explained by the fact that a finite 
kinetic energy (2.1 $\mu K$) was used
in the close-coupled calculation whereas a zero kinetic energy was used in the perturbative 
calculation. We note that an increase in the kinetic energy by an order of magnitude decreases 
the non-perturbative line shift fit parameter by approximately 2\%.

At higher intensities the perturbative results remain a fair approximation to
the non-perturbative results, however small non-linear differences are evident,
affecting the line shifts by up to 5\%. Additionally, a maximum intensity is
found that limits the visibility of the resonances. This laser saturation
should be observable in the laboratory and suggests that ultracold
photoassociation to shallow potentials is only useful at lower laser
intensities. Although not reported here, it can be shown \cite{Portier06} that
the perturbative line width is also linearly dependent upon laser intensity.
This is obviously not true in the present non-perturbative calculation, 
where a significant quadratic behavior is observed.

We conclude by noting that our theoretical findings are consistent with the limited experimental
data available \cite{Kim05}.  Our values for the shift ratios $s_{1}(v=1)/s_{1}(v=0)$ and 
$s_{1}(v=2)/s_{1}(v=0)$ of 1.81 and 4.57 for $\sigma^{-}$ polarization  lie within the 
experimental values of $1.71 \pm 0.14$ and $4.20 \pm 0.48$ respectively, and our $v=0$ widths of 
3.24 MHz at 9 mW/cm$^{2}$ and 11.9 MHz at 5 W/cm$^{2}$ are comparable to the experimental 
values estimated from Fig. 2 of \cite{Kim05} of 3 MHz and 10 MHz respectively.

\begin{acknowledgments}
IBW would like to thank A. S. Dickinson for helpful discussions and the suggestion that
we investigate the use of a modified radiative coupling.
\end{acknowledgments}

\appendix

\section{The $|j_{1}j_{2}j\Omega_{j},Jm_{J}p\rangle$ Basis}

In the $jj$ coupling scheme, the total angular momenta $\bm{j}_{i}$
of each atom are coupled to form the total electronic angular momentum 
$\bm{j}=\bm{j}_{1}+\bm{j}_{2}$ which is then coupled to the relative 
angular momentum $\bm{l}$ of the nuclei to form the total angular
momentum $\bm{J}=\bm{j}+\bm{l}$. This gives the basis states in the 
space-fixed reference frame
\begin{equation}
\label{cw-A1}
|\gamma j_{1}j_{2}jlJm_{J}\rangle=\sum_{m_{j}m_{l}}C_{m_{j}m_{l}m_{J}}^{jlJ}
|\gamma j_{1}j_{2}jm_{j}\rangle|lm_{l}\rangle
\end{equation}
where $|\gamma j_{1} j_{2} j m_{j} \rangle $ are the eigenstates of the 
two asymptotically free atoms and $|l m_{l} \rangle =
Y_{l,m_{l}}(\theta ,\phi)$
are the relative motion eigenstates.
Here $\gamma=(\gamma_{1},\gamma_{2})$ and $\gamma_{i}=
\{\bar{\gamma}_{i},L_{i},S_{i}\}$ where $\bm{L}_{i}$ and 
$\bm{S}_{i}$ are
the total orbital and total spin angular momentum respectively of the 
individual atoms. The label $\bar{\gamma}_{i}$ denotes any additional 
quantum numbers needed, including those specifying the electron configurations.
All subscripted $m$ quantities denote projections on the $Oz$ axis of the 
space-fixed frame.

The transformation from the space-fixed frame to the molecular frame is
\begin{equation}
\label{cw-A2}
|\gamma j_{1}j_{2}jm_{j}\rangle=\sum_{\Omega_{j}}D_{m_{j}\Omega_{j}}^{j*}
(\phi,\theta,0)|\gamma j_{1}j_{2}j\Omega_{j}\rangle
\end{equation}
where subscripted $\Omega$ quantities indicate projections along the 
intermolecular axis and $D^j_{m_j\Omega_j}$ is the Wigner rotation matrix
\cite{Brink68}. Expressing the relative motion state $|lm_{l} \rangle $ as 
a rotation matrix gives
\begin{eqnarray}
\label{cw-A3}
|\gamma j_{1}j_{2}jlJm_{J}\rangle & = & \sum_{m_{j}m_{l}\Omega_{j}}
C_{m_{j}m_{l}m_{J}}^{jlJ}\sqrt{\frac{2l+1}{4\pi}}
D_{m_{l}0}^{l*}(\phi,\theta,0)  \nonumber \\
& \times & D_{m_{j}\Omega_{j}}^{j*}(\phi,\theta,0)
|\gamma j_{1}j_{2}j\Omega_{j}\rangle .
\end{eqnarray}
Combining the rotation matrices and using the sum rule for Clebsch-Gordan 
coefficients reduces (\ref{cw-A3}) to
\begin{eqnarray}
\label{cw-A5}
|\gamma j_{1}j_{2}jlJm_{J}\rangle & = & \sum_{\Omega_{j}}(-1)^{j-\Omega_{j}}
C_{\Omega_{j}-\Omega_{j}0}^{jJl}   \nonumber \\
& \times &  N_{m_{J}\Omega_{j}}^{J}(\theta,\phi)|\gamma j_{1}j_{2}j\Omega_{j}\rangle 
\end{eqnarray}
where $N_{m_{J}\Omega_{j}}^{J}$ is the symmetric top function defined as
\begin{equation}
\label{cw-A6}
N_{m_{J}\Omega_{J}}^{J} \equiv \sqrt{\frac{2J+1}{4\pi}} 
D_{m_{J}\Omega_{J}}^{J*}(\phi,\theta,0).
\end{equation}
Equation (\ref{cw-A5}) can be interpreted as a coupling of $j$
and $J$ to result in $l$ and naturally introduces the basis states
\begin{equation}
\label{cw-A7}
|\gamma j_{1}j_{2}j\Omega_{j}Jm_{J}\rangle \equiv  N_{m_{J}
\Omega_{j}}^{J}|\gamma j_{1}j_{2}j\Omega_{j}\rangle .
\end{equation}

The derivation of \eqref{cw-A7} has not taken into account any of the
symmetry requirements of the system. Following \cite{Nikitin84}, we define 
symmetric states that are eigenstates of the operator $\hat{I}$, the 
inversion operator of the total wavefunction through the centre of 
charge of the molecule. The eigenvalues of $\hat{I}$ are $(-1)^w$ with $w=0$ for
\textit{gerade} symmetry and $w=1$ for \textit{ungerade} symmetry. For $\Omega_{j}=0$ states,
we can also identify the quantum number of $\hat{\sigma}_{v}$, the
reflection operator of the total wavefunction through a plane containing
the intermolecular axis. For identical nuclei the symmetric states for 
$jj$ and $LS$ couplings are
\begin{eqnarray}
|\gamma_{1}j_{1}\gamma_{2}j_{2}j\Omega_{j}w\rangle & = & N_{jj}
(|\gamma_{1}j_{1}\gamma_{2}j_{2}j\Omega_{j}\rangle  \nonumber \\
& + & (-1)^{p_{jj}}|\gamma_{2}j_{2}\gamma_{1}j_{1}j\Omega_{j}\rangle) \label{cw-A8} 
\end{eqnarray}
and
\begin{eqnarray}
|\gamma_{1}\gamma_{2}LS\Omega_{L}\Omega_{S}w\rangle & = & N_{LS}
(|\gamma_{1}\gamma_{2}LS\Omega_{L}\Omega_{S}\rangle  \nonumber\\
& + & (-1)^{p_{LS}}|\gamma_{2}\gamma_{1}LS\Omega_{L}\Omega_{S}\rangle) \label{cw-A9}
\end{eqnarray}
where the explicit ordering of $\gamma_{1}$ and $\gamma_{2}$ indicates
the order in which the angular momenta are coupled and $p$, which labels the symmetry under 
permutation of the labels $1 \leftrightarrow 2$, is uniquely related to $w$.
For the  $jj$ state (\ref{cw-A8}), $p_{jj}=w+w_{1}+w_{2}+N+j_{1}+j_{2}-j$
where $w_{i}$ is the symmetry under inversion of the electronic wavefunction 
of atom $i$ about its nucleus and $N$ is the total number of electrons per
atom.  The $LS$ state (\ref{cw-A9}) has $p_{LS}=w+w_{1}+w_{2}+N+S_{1}+S_{2}-S
+L_{1}+L_{2}-L$.

The normalization constants $N_{jj}$ and $N_{LS}$ are  
\begin{equation}
\label{cw-A10}
N_{jj}= \left\{ \begin{array}{ll}
	\frac{1}{\sqrt{2}} & \text{for }(\gamma_{1},j_{1})\neq (\gamma_{2},j_{2})\\
	\frac{1}{2} & \text{for }(\gamma_{1},j_{1})=(\gamma_{2},j_{2})
        \end{array} \right.
\end{equation}
and
\begin{equation}
\label{cw-A11}
N_{LS}= \left\{ \begin{array}{ll}
	\frac{1}{\sqrt{2}} & \text{for }\gamma_{1}\neq\gamma_{2}\\
        \frac{1}{2} & \text{for }\gamma_{1}=\gamma_{2}
        \end{array}. \right.
\end{equation}

The transformation between the two bases (\ref{cw-A8}) and (\ref{cw-A9}) is 
\begin{eqnarray}
\label{cw-A12}
|\gamma_{1}j_{1}\gamma_{2}j_{2}j\Omega_{j}w\rangle & = & 
\sum_{LS\Omega_{L}\Omega_{S}}\frac{N_{jj}}{N_{LS}}
F_{LS\Omega_{L}\Omega_{S}}^{j_{1}j_{2}j\Omega_{j}} \nonumber \\
& \times & |\gamma_{1}\gamma_{2}LS\Omega_{L}\Omega_{S}w\rangle
\end{eqnarray}
where
\begin{eqnarray}
\label{cw-A13}
F_{LS\Omega_{L}\Omega_{S}}^{j_{1}j_{2}j\Omega_{j}} &= & 
[(2S+1)(2L+1)(2j_{1}+1)(2j_{2}+1)]^{\frac{1}{2}} \nonumber \\
& \times & C_{m_{L}m_{S}m_{j}}^{LSj} \left\{
	\begin{array}{ccc}
	L_{1} & L_{2} & L\\
	S_{1} & S_{2} & S\\
	j_{1} & j_{2} & j
	\end{array}
\right\}
\end{eqnarray}
where the $\{\cdots \! \}$ is the Wigner $9-j$ symbol and the implicit set of quantum
numbers $(\gamma_1,\gamma_2)$ has been suppressed.

\section{Matrix Elements}

\subsection{Kinetic terms}

The radial kinetic term has the form
\begin{eqnarray}
\label{cw-B2}
\langle a^{\prime}|\hat{T}\frac{1}{R}G_{a}(R)|a\rangle & = & -\frac{\hbar^{2}}{2\mu R}
\left(\frac{d^{2}G_{a}}{dR^{2}}\delta_{a a^{\prime}} 
+2\frac{dG_{a}}{dR}\langle a^{\prime}|\frac{d|a\rangle}{dR}\right. \nonumber \\
& + & \left. G_{a}\langle a^{\prime}|\frac{d^{2}|a\rangle}{dR^{2}}\right)
\end{eqnarray}
where $|a \rangle \equiv |\Phi_{a}(R,q)\rangle $ and $a$ represents the quantum 
numbers $\{\gamma_{1},\gamma_{2},j_{1},j_{2},j,\Omega_{j},w,J,m_{J}\}$. 
As the basis states are assumed to vary little with respect to $R$, the last two terms 
in (\ref{cw-B2}) are negligible at the long ranges considered in this investigation.  

In this investigation the adiabatic excited state $|\psi_e^0\rangle$ is 
a combination of these basis states (see (\ref{cw-20})) and the coefficients 
$C_{ea}(R)$ do vary considerably with $R$.
If we assume no other excited states are coupled to the system 
(i.e. the adiabatic approximation is exact) then we only require the radial term 
(\ref{cw-B2}) with $| a \rangle = |a^{\prime} \rangle = |\psi^{0}_{e}\rangle $.
The second term of (\ref{cw-B2}) is then zero as
\begin{equation}
\label{cw-B3}
\langle \psi_e^0|\frac{d|\psi_e^0\rangle}{dR}=\frac{1}{2}\frac{d}{dR}\;\sum_{a}
|C_{ea}(R)|^2 = 0.
\end{equation}
The third term, however, is non-zero and gives rise to the kinetic correction term
\begin{equation}
\label{cw-B3a}
V^\text{KC}_e = -\frac{\hbar^2}{2\mu R} \sum_{a} C_{ea}(R) \frac{d^2 C_{ea}(R)}{dR^2}.
\end{equation}

The matrix elements of the rotational kinetic term 
\begin{equation}
\label{cw-B4}
\hat{H}_{\text{rot}}=\frac{\hat{l}^{2}}{2\mu R^{2}}
\end{equation}
are evaluated using the expansion of $\hat{l}^2$ in terms of ladder operators
\begin{equation}
\hat{l}^2 = \hat{J}^2 + \hat{j}^2 - (2\hat{J}_z\hat{j}_z + \hat{J}_+\hat{j}_- +
\hat{J}_-\hat{j}_+)
\end{equation}
where the subscripts refer to molecule-fixed axes, $\hat{J}_\pm \equiv \hat{J}_x
\pm i\hat{J}_y$, and $\hat{j}_\pm \equiv \hat{j}_x \pm i\hat{j}_y$. The action of
$\hat{J}_\pm$ is irregular \cite{Vleck51} due to the rotation of $N^J_{m_J \Omega_j}$
and is given by
\begin{equation}
\hat{J}_\pm N^J_{m_J \Omega_j} = \hbar \sqrt{J(J+1) - \Omega_j(\Omega_j \mp 1)}
N^J_{m_J \Omega_j\mp 1}.
\end{equation}

Hence the matrix
elements of $\hat{l}^2$ are
\begin{eqnarray}
\label{cw-B5}
\langle a^{\prime}|\hat{l^{2}}|a\rangle & = & \hbar^{2} \delta_{\rho
\rho^{\prime}}\left\{
\left[J(J+1)+j(j+1)-2\Omega_{j}^{2}\right]\delta_{\Omega_{j}^{\prime}\Omega_{j}}
\right. 
\nonumber \\
 & - & \left. K^{-}_{Jj\Omega_{j}}\delta_{\Omega_{j}^{\prime},\Omega_{j}-1} - 
K^{+}_{Jj\Omega_{j}}\delta_{\Omega_{j}^{\prime},\Omega_{j}+1} \right\} 
\end{eqnarray}
where $|a \rangle \equiv |\Phi_{a}(R,q)\rangle $, $a$ represents the quantum 
numbers $\{\gamma_{1},\gamma_{2},j_{1},j_{2},j,\Omega_{j},w,J,m_{J}\}$,
$\rho$ denotes the set of quantum numbers $\{\gamma_1,\gamma_2,j_1,j_2,j,w,J,m_J\}$ and
\begin{eqnarray}
\label{cw-B6}
K^{\pm}_{Jj\Omega_{j}} &= &\left[J(J+1) - \Omega_j(\Omega_j \pm 1)\right]^{\frac{1}{2}} 
\nonumber \\ 
& \times & \left[ j(j+1) - \Omega_j (\Omega_j \pm 1)\right]^{\frac{1}{2}}
\end{eqnarray}
The terms non-diagonal in $\Omega_{j}$ are called the Coriolis couplings and 
are often negligible. This is the case for purely long-range bound states in
metastable helium \cite{Venturi03}.

\subsection{Electronic term}

We wish to express the matrix elements of $\hat{H}_{\text{el}}$ in terms the
Born-Oppenheimer potentials ${}^{2S+1}\Lambda^{\pm}_{w}(R)$ defined by
the eigenvalue equation
\begin{equation}
\label{cw-B7a}
\hat{H}_{\text{el}}|LS\Omega_{L}\Omega_{S}w\rangle =  
[{}^{2S+1}\Lambda^{\pm}_{w}(R)+ E^{\infty} ]|LS\Omega_{L}\Omega_{S}w\rangle
\end{equation}
where $\Lambda \equiv|\Omega_{L}|$ and $E^{\infty}$ is the asymptotic energy of 
the state. 

The matrix elements in the basis (\ref{cw-A8}) are evaluated by transforming to 
the basis (\ref{cw-A9}) using (\ref{cw-A12}) and then applying
(\ref{cw-B7a}) to obtain 
\begin{eqnarray}
\label{cw-B7}
\langle a^{\prime}|\hat{H}_{\text{el}}|a\rangle & = & \delta_{\eta \eta^{\prime}}
\;\sum_{LS\Omega_{L}\Omega_{S}} \frac{N_{jj}^2}{N_{LS}^2} 
F_{LS\Omega_{L}\Omega_{S}}^{j_{1}^{\prime}j_{2}^{\prime}j^{\prime}\Omega_{j}}
\nonumber  \\
& \times & [{}^{2S+1}\Lambda^{\pm}_{w}(R)+E_{a}^{\infty}]
F_{LS\Omega_{L}\Omega_{S}}^{j_{1}j_{2}j\Omega_{j}}
\end{eqnarray}
where $\eta =\{\gamma_{1},\gamma_{2},\Omega_{j},w,J,m_{J}\}$.
For our case of metastable helium involving the $2s2s$ and $2s2p$ states, $N_{jj} =
N_{LS}$.

\subsection{Fine-structure term}

The total fine structure term $\hat{H}_{\text{fs}}$ is the sum of the 
fine structure terms for the individual atoms and is diagonal in the 
basis (\ref{cw-A7}) in the asymptotic region of free atoms.
For the long-range molecular states considered in this work, the total atomic 
angular momenta are considered to be approximately good quantum numbers. Hence the
matrix elements of the total fine-structure term are
\begin{eqnarray}
\label{cw-B9}
\langle a^{\prime}|\hat{H}_{\text{fs}}|a\rangle & = & \langle a^{\prime}|
\hat{H}_{\text{fs}}^{1}+\hat{H}_{\text{fs}}^{2}|a \rangle \nonumber \\
 & = & \delta_{a a^{\prime}}(\Delta E_{\gamma_{1}j_{1}}^{\text{fs}}+
\Delta E_{\gamma_{2}j_{2}}^{\text{fs}})
\end{eqnarray}
where $\hat{H}_{\text{fs}}^{i}$ represents the fine-structure interaction
and $\Delta E_{\gamma_{i}j_{i}}^{\text{fs}}$
the fine-structure splittings for atom $i$.

\subsection{Laser interaction term}

For radiation of a given circular polarization $\bm{\epsilon}_{\lambda}$ in the 
space-fixed frame, where $\lambda=0, \pm 1$ for $\pi , \sigma ^{\pm}$ polarization, 
the laser-matter interaction (\ref{cw-9}) can be expanded in a spherical basis using
\begin{equation}
\label{cw-B10}
\bm{\epsilon}_{\lambda} \cdot \bm{d}=  \sum_{\xi=0,\pm 1} (-1)^{\xi}
( \bm{\epsilon}_{\lambda})_{-\xi} d_{\xi}
\end{equation}
where $(\bm{\epsilon}_{\lambda})_{-\xi} = \delta_{\lambda,\xi}$. 
The matrix elements of 
$\hat{H}_{\text{int}}$  are, after rotation to the molecular frame and transformation to the 
$LS$ basis states, 
\begin{eqnarray}
\label{cw-B11}
\langle a^{\prime}|\hat{H}_{\text{int}}|a\rangle & = & A_{\lambda}  \frac{N_{jj}^{\prime}}
{N_{LS}^{\prime}} \frac{N_{jj}}{N_{LS}} \;
\sum_{L^{\prime}S^{\prime}\Omega_L^{\prime}\Omega_S^{\prime}} 
\sum_{LS\Omega_L\Omega_S}  
F^{j_1^{\prime}j_2^{\prime}j^{\prime}\Omega_j^{\prime}}_
{L^{\prime}S^{\prime}\Omega_L^{\prime}\Omega_S^{\prime}} 
\nonumber \\
& \times & F^{j_1j_2j\Omega_j}_{LS\Omega_L\Omega_S} \int\! \sin\theta \, d\theta \, d\phi \; 
N^{J^{\prime}}_{m_{J^{\prime}}\Omega_j^{\prime}} 
D^{1*}_{\lambda b} N^{J}_{m_J\Omega_j}  \nonumber \\
& \times & \langle \gamma ^{\prime}L^{\prime}S^{\prime}\Omega_L^{\prime}
\Omega_S^{\prime}w^{\prime}| d_b | \gamma LS\Omega_L\Omega_S w \rangle 
\end{eqnarray}
where $A_{\lambda} =(-1)^{\lambda}\sqrt{\frac{I}{2\epsilon_{0}c}}$, 
$D^j_{m_j\Omega_j} \equiv D^j_{m_j\Omega_j}(\theta,\phi,0)$, and $b$ labels the
spherical basis components in the molecular frame.

The terms involving $J$ and $J^{\prime}$ can be expanded and the integration over the 
interatomic polar coordinates performed, to give
\begin{gather}
\label{cw-B12}
\sqrt{\frac{(2J^{\prime}+1)(2J+1)}{4\pi}} \int \! \sin\theta \, d\theta \, d\phi\,
 D^{J^{\prime}}_{m_J^{\prime}\Omega_j^{\prime}} D^{1*}_{\lambda b} D^{J*}_{m_J\Omega_j} 
\nonumber \\
= \sqrt{\frac{2J+1}{2J^{\prime}+1}} C^{J1J^{\prime}}_{m_J \lambda m_J^{\prime}} 
C^{J1J^{\prime}}_{\Omega_j b \Omega_j^{\prime}}.
\end{gather}
The matrix element of $d_b=d^{1}_{b}+d^{2}_{b}$ between $LS$ states must be evaluated 
under proper symmetry considerations \cite{Burke99}. For the helium 2$s$-2$p(0^+_u$) 
transition this results in
\begin{gather}
\label{cw-B13}
\langle L^{\prime}(=1)S^{\prime}\Omega_L^{\prime}\Omega_S^{\prime}w^{\prime}(=1)|d_b|
L(=0)S \Omega_L(=0)\Omega_S w \rangle  \nonumber \\
 = \delta_{SS^{\prime}}\delta_{\Omega_{S}\Omega_{S}^{\prime}}
\delta_{b\Omega_L^{\prime}}\frac{d_{\text{at}}^{\text{sp}}} 
{\sqrt{2}}\left[1+(-1)^{1+S+w^{\prime}}\right]
\end{gather}
where $d_{\text{at}}^{\text{sp}}$ is the reduced matrix element of the dipole operator 
between the 2$s$ and 2$p$ atomic states. Only gerade ($w=0$) ground states are coupled to 
the excited state, and because metastable states must satisfy 
$(-1)^{S+w}=1$ due to their bosonic nature, only ${}^1\Sigma^+_g$ and ${}^5\Sigma^+_g$ 
states are coupled to the excited state.

After simplification the complete matrix element becomes
\begin{eqnarray}
\label{cw-B14}
\langle a^{\prime}| \hat{H}_{\text{int}} | a \rangle & = & (-1)^{\lambda} \sqrt{\frac{I}{\epsilon_{0}c}}
\; \sqrt{\frac{2J+1}{2J^{\prime}+1}}\; F^{j_1^{\prime}j_2^{\prime}j^{\prime}0}_ {1,j,-\Omega_j,\Omega_j} 
\nonumber \\
& \times & C^{J1J^{\prime}}_{\Omega_j, -\Omega_j, 0} C^{J1J^{\prime}}_{m_J,\lambda,m_J^{\prime}} 
d_{\text{at}}^{\text{sp}} 
\end{eqnarray}
assuming that $w^{\prime} = 1$ and $w = 0$.

\end{document}